\documentclass[%
reprint,
superscriptaddress,
nofootinbib,
amsmath,
amssymb,
aps,
prd,
floatfix,
showkeys,
]{revtex4-2}

\usepackage{graphicx}% Include figure files
\usepackage{dcolumn}% Align table columns on decimal point
\usepackage{bm}
\usepackage{subfigure}

\usepackage{epsfig}
\usepackage{mathtools}
\usepackage{booktabs}
\usepackage{aas_macros}

\usepackage{natbib}
\usepackage[colorlinks]{hyperref}
\hypersetup{linkcolor=magenta,citecolor=blue}
\usepackage[colorlinks]{hyperref}

\usepackage[normalem]{ulem}
\usepackage{fancyvrb}
\usepackage{fancyhdr}
\newcommand{\beq}{\begin{equation}}
\newcommand{\eeq}{\vspace{0cm} \end{equation}}
\newcommand{\beqq}{\setlength\arraycolsep{2pt}\begin{eqnarray}}
\newcommand{\eeqq}{\vspace{0cm} \end{eqnarray}}

\begin{document}

\title{Impact of dark energy on the structure of neutron stars: The vacuum case}

% Interplay of Dark Energy and Standard Model Matter in Neutron Stars: An Admixed Perspective

\author{L. F. Araújo}
\email{loreanyfa@usp.br}
\affiliation{Departamento de Astronomia, Universidade de S\~{a}o Paulo \\ Rua do Mat\~{a}o, 1226 - 05508-900, S\~ao Paulo, SP, Brazil}

\author{J. A. S. Lima}
\email{jas.lima@iag.usp.br}
\affiliation{Departamento de Astronomia, Universidade de S\~{a}o Paulo \\ Rua do Mat\~{a}o, 1226 - 05508-900, S\~ao Paulo, SP, Brazil}

\author{G. Lugones}
\email{german.lugones@ufabc.edu.br}
\affiliation{Universidade Federal do ABC, Centro de Ci\^encias Naturais e Humanas, Avenida dos Estados 5001- Bang\'u, CEP 09210-580, Santo Andr\'e, SP, Brazil.}


\begin{abstract}
The potential role of a cosmic vacuum dark component in the properties of neutron stars is investigated. It is assumed that the static, spherically symmetric distribution of matter within neutron stars is supported by two distinct components: ordinary matter and a vacuum fluid.  For normal matter we use a set of state-of-the-art nuclear matter equations of state, each grounded in  nuclear physics experiments. The vacuum energy component is inhomogeneously distributed  within the star and obeys the standard equation of state ($p_v = -\epsilon_v$). This is characterized by an energy density fraction $y = \epsilon_m/ (\epsilon_m + \epsilon_v)$, which we model as either a constant or radius-dependent. Our findings reveal that the inclusion of vacuum energy significantly affects the mass-radius relationships in neutron stars, influencing both the maximum achievable masses and the qualitative form of these relationships. Some constraints from current multimessenger observational data limiting the amount of vacuum energy within neutron stars are also discussed.
\end{abstract}


\maketitle
\bigskip

\section{Introduction}

Numerous astrophysical and cosmological studies suggest that most of the observed Universe is made up of dark matter and dark energy, both of which have unknown origins and compositions. The clustering of dark matter helps explain the observed rotation of galaxies and the Universe's large-scale structure, while a uniform dark energy component is required to account for the Universe's late-time acceleration. The standard explanation for dark matter assumes the existence of yet unknown particles weakly coupled to the Standard Model.  Meanwhile, it remains uncertain whether dark energy represents the manifestation of a novel field, a challenge to General Relativity thereby leading to alternative gravity theories, or an entirely different phenomenon.  

While the role of dark energy at large scales has been extensively discussed, our understanding of its effects at more local scales, particularly at the level of individual astronomical objects such as neutron stars, remains much less explored. Over the years, some models have been proposed to understand the presence of dark energy in compact objects \cite{Beltracchi:2018ait}. In the 1960s, Gliner put forward the idea of objects with a $p=-\epsilon$ equation of state (EOS) at their centers \cite{gliner1966algebraic}. The Bardeen spacetime, proposed as the first concrete solution, represented a nonsingular, asymptotically flat, spherically symmetric spacetime with zero, one, or two event horizons, depending on a specific parameter \cite{Bardeen1968,Borde:1994ai,Borde:1996df,Zhou:2011aa}. In the 1980s, the formation and gravitational effects of false vacuum bubbles in true vacuum (and vice versa) were studied, with some analysis suggesting their potential role in wormhole formation and localized inflation \cite{Coleman:1980aw, Sato:1981bf, Blau:1986cw, Farhi:1986ty}. The 2000s  witnessed a resurgence of interest in the $p=-\epsilon$ concept. Chapline et al. proposed ``dark energy stars'', gravitationally stable compact objects with a dark energy core and a microscopic quantum critical layer in place of an event horizon \cite{Chapline:2004jfp}. Another model, the stiff shell gravastar, was introduced by Mazur and Mottola, characterized by a positive pressure stiff matter surface layer connected to the dark energy core and external vacuum by junction layers \cite{Mazur:2001fv,Mazur:2004fk}. Research has extensively examined these models for their potential behaviors and stability \cite{Visser:2003ge, Chan:2011ayt, DeBenedictis:2005vp, Carter:2005pi}. Additional models, including non-singular black holes \cite{Dymnikova:1992ux}, have also been explored to understand dark energy in compact objects, as well as in the gravitational collapse in its variant  associated with the vacuum energy density, minimally-coupled scalar fields, and in anisotropic regimes \cite{Campos_2012, smerechynskyi2021impact, sagar2022hybrid, rej2023charged, das2023dark}.

In contrast to the more theoretical exploration of dark energy compact objects, there exists a significant body of work in the literature detailing the possible role of dark matter within realistic astrophysical objects.  Numerous theoretical investigations have proposed that dark matter can become part of neutron stars either through direct gravitational capture or through non-gravitational interactions, resulting in so-called ``admixed'' stars that contain a mixture of standard model matter and dark matter \cite{Li:2012qf,Mukhopadhyay:2015xhs,Baym:2018ljz,Ellis:2018bkr,Dengler:2021qcq,Kain:2021hpk,Sen:2021wev,Jimenez:2021nmr,Miao:2022rqj}.  In these admixed neutron stars, it is suggested that dark matter may exist as a small core \cite{Jimenez:2021nmr}, or as a halo around the star that does not affect its visible radius \cite{Miao:2022rqj}. In some extreme cases, neutron stars might be almost entirely composed of dark matter, creating highly compact dark objects with very low masses \cite{Horowitz:2019aim}.  Several studies have also contemplated the existence of neutron stars composed of particles from a hidden dark-QCD sector, asymmetric dark matter fermions, or asymmetric bosons \cite{Maedan:2019mgz,Narain:2006kx,Kouvaris:2015rea,Gresham:2018rqo,Leung:2022wcf,Rutherford:2022xeb}. 
Many observational consequences have been hypothesized as a result of the existence of dark matter within neutron stars. It has been suggested that dark matter could influence the mass-radius relation~\cite{Panotopoulos:2017idn}, affect the tidal deformability \cite{Nelson:2018xtr}, and potentially modify the gravitational wave signal emitted during a binary neutron star merger \cite{Ellis:2017jgp}.

In parallel with the gravitational capture of dark matter by neutron stars, the potential influence of dark energy in compact objects, given their strong gravitational fields, is an intriguing hypothesis that merits further exploration.  While prior studies have essentially addressed gravitationally stable compact entities composed partially or entirely of pure dark energy, the coexistence of dark energy with standard model matter in such contexts remains unexplored. Our investigation adopts a perspective akin to ``admixed" dark matter stars, proposing the interweaving of dark energy and ordinary matter within the star. This ``admixed" framework gains enhanced astrophysical plausibility, representing a more feasible scenario. 

Among several possibilities for describing the dark energy candidates, the simplest and most compelling one is the cosmological constant, $\Lambda$. In the context of the Einstein field equations, the $\Lambda$-term behaves like an additional energy component with a constant energy density, $\epsilon_v = \Lambda/8\pi G$, and pressure, $p_v = -\epsilon_v$. When combined with the cold dark matter (CDM) component, the resulting model is commonly referred to as the cosmic concordance cosmology ($\Lambda$CDM). Over the last two decades, it has become the standard cosmological model. However, the $\Lambda$CDM cosmology is currently facing observational challenges, particularly from the so-called Supernovae-CMB tension \cite{riess2019}. Briefly, estimates of the Hubble constant based on supernovae observations from the SHOES collaboration, measurements at intermediate redshifts from various tests \cite{cunha2007}, and observations of strong lensing time-delays suggest a best-fit value of $H_0 = 74 \, \mathrm{km/s/Mpc}$, whereas the latest Planck CMB data favors a value closer to $68 \, \mathrm{km/s/Mpc}$. This indicates a discrepancy of around $5.6\sigma$ (see, e.g., \cite{abdalla2022cosmology} for a recent review discussing the main problems with the $\Lambda$CDM model).

The cosmological constant and coincidence problems have motivated the idea that $\Lambda$ in the homogeneous and isotropic case would be a time-dependent quantity, i.e., $\Lambda = \Lambda(t)$ due to its interaction with all the matter fields in the Universe. In particular, it has been shown that $\dot{\Lambda}(t) < 0$, suggesting that $\Lambda$ is small today because the Universe is too old \cite{carvalho1992}. This result is also required by the second law of thermodynamics because entropy growth ($\dot{S} > 0$) occurs only when the vacuum energy density decreases during expansion \cite{Lima1996,lima2004}. In principle, this explains the current low value of the vacuum energy density and, potentially, the coincidence problem \cite{Lima_2016}. Naturally, in an inhomogeneous static case, such as in compact objects, the cosmological constant would be a function only of position, $\Lambda = \Lambda(r)$.

In this context, prompted by cosmology, it is of interest to investigate the implications of spatially varying vacuum energy density on compact objects such as neutron stars. In cosmology, we have seen that the strength of $\Lambda$ is greater at early times when the energy density of the Universe is very high. Similarly, one may conjecture that for neutron stars, the vacuum energy density diminishes from their core to their surface, as occurs for the pressure within the star itself.
 
On the other hand, recent developments in multi-messenger astronomy have significantly advanced our understanding of neutron star properties. Pulsars such as PSR J1614-2230 \citep{Demorest:2010ats, Arzoumanian:2018tny} and PSR J0348+0432 \citep{Antoniadis:2013amp}, each with masses around $2 M_\odot$, have provided crucial data for studying neutron star matter. The subsequent discovery of PSR J0740+6620 \citep{Cromartie:2020rsd,Fonseca:2021rfa}, another high-mass pulsar, reinforces the need for equations of state to be capable of accommodating neutron stars with masses exceeding $2 M_\odot$. Collaborative efforts between the NICER telescope and XMM-Newton have yielded valuable constraints on the mass and radius of pulsars like PSR J0740+6620 \citep{Miller:2021tro,Riley:2021anv} and PSR J0030+0451 \citep{Miller:2019pjm,Riley2019anv}. Additionally, the LIGO-Virgo gravitational wave observatory has contributed crucial data through neutron star merger events, such as GW170817 \citep{LIGOScientific:2017vwq, Abbott:2018exr}, providing insights into neutron star masses, radii, and tidal deformabilities.  These combined observations offer opportunities to constrain the presence of vacuum energy within compact objects, contributing to our understanding of fundamental physics in astrophysical contexts.

This paper is structured as follows. Sec.~\ref{sec:structure} details the derivation of the equations governing stellar structure, specifically incorporating the effects of vacuum energy into these equations. Our study examines two different models. In Sec.~\ref{sec:yconst}, we consider a simpler model where the ratio of the energy densities of ordinary matter to vacuum energy is treated as a constant, acting as a free parameter. Sec.~\ref{yvar} extends this approach to include the derivation of structure equations that allow for density-dependent variations in this ratio. In Section~\ref{sec:EOS}, we describe a set of state-of-the-art nuclear EOS used to represent normal matter in neutron stars. Sec.~\ref{sec:results} presents the outcomes of our calculations, with a particular focus on the mass-radius relationship and tidal deformabilities of neutron stars under these models. Finally, in Sec.~\ref{sec:conclusions}, we summarize our findings and discuss their implications for understanding the potential role of dark energy in neutron star physics.

\section{Stellar structure equations with a vacuum component}
\label{sec:structure}

In this section, we will incorporate the vacuum energy into the composition of neutron stars, assuming that it is admixed with ordinary matter. To this end, we will consider these compact stars to be spherically symmetric, static, and isotropic. Consequently, we can express the metric of these objects as:
\begin{equation}
ds^2 = e^{2\nu(r)} dt^2 - e^{2\mu(r)} dr^2 - r^2 (d \theta^2 + \sin^2 \theta d \phi^2) .
\label{metricsch}
\end{equation}
It is common practice to introduce a new metric function $m(r)$ defined by $e^{2\mu(r)} = 1 - \frac{2m(r)}{r}$. The quantity $m(r)$ carries the interpretation of ``mass enclosed within a radius $r$'', and, from now on, units such that $c = G = 1$ are adopted.

It will be assumed here that the vacuum component ($v$) and ordinary matter ($m$) are characterized as separated perfect fluid components interacting only gravitationally. The total energy-momentum tensor, which encompasses both vacuum energy and ordinary matter, can be expressed as:
\begin{equation}
T^{(t)}_{\mu \nu} = T^{(m)}_{\mu \nu} + T^{(v)}_{\mu \nu},
\end{equation}
where the individual components have the following form:
\begin{equation}
T_{\mu \nu} = (p + \epsilon) u_{\mu} u_{\nu} - p g_{\mu\nu},
\end{equation}
being $p$ and $\epsilon$ the pressure and energy density, respectively.

As one may check, for a comoving observer ($u^{\mu} = (\sqrt g_{00})^{-1}\delta^{\mu}_0$,  $u_{\mu}=\sqrt g_{00} \delta_{\mu 0}$), the Tolman-Oppenheimer-Volkoff (TOV) equations describing the structure of stellar objects take the following form:
\begin{eqnarray}
\frac{dm}{dr} &=& 4 \pi \epsilon r^2, \label{mass} \\
\frac{dp}{dr} &= &-\frac{(p + \epsilon) \left [m + 4\pi r^3 p\right]}{r^2 - 2mr}\label{eqh}, \\
\frac{d \nu}{dr} & =& \frac{ m + 4\pi r^3 p}{r^2 - 2mr},
\label{nueq}
\end{eqnarray}
where  $m$, $p$ and $\epsilon$ are the sum of contributions from ordinary matter and vacuum:  
\begin{eqnarray}
m  &= & m_m + m_v,    \\
p &= & p_m + p_{v},  \\
\epsilon & = & \epsilon_m + \epsilon_v. 
\end{eqnarray}

To solve the TOV equations, it is necessary to supplement them with equations of state (EOSs) that describe the relationship between pressure and energy density within the stellar interior. In the current scenario involving both ordinary matter and vacuum energy, we will specify separate EOSs for each component. For vacuum energy, we will adopt the standard relation $p_{v}=-\epsilon_{v}$. 
Regarding ordinary matter, we will use a set of state-of-the-art EOSs for hadronic matter. These EOSs will be thoroughly described in Section \ref{sec:EOS}. All of these hadronic EOSs are in agreement with current astrophysical constraints on neutron stars and meet experimental restrictions at the nuclear saturation density, as detailed in Sec. \ref{sec:EOS}.

Once the EOSs for ordinary matter and vacuum energy are specified, the three TOV equations will contain four unknown functions that need to be determined: $m(r)$, $p_m(r)$, $p_v(r)$, and $\nu(r)$.
To close the system and ensure a complete solution, an additional condition is required. In order to simplify the analysis, we adopt the following approach: we assume that ordinary matter constitutes a fraction $y$ of the total energy density, while vacuum energy constitutes a fraction $1-y$.  The fraction $y$ will be specified in different ways to investigate the impact of varying vacuum energy contents on the stellar structure. In Sec. \ref{sec:yconst}, we will assign $y$ a set of constant values. Alternatively, in Sec. \ref{yvar}, we will allow $y$ to vary across different layers of the star. 
Mathematically, $y$ can be expressed as:
\begin{equation}
y = \frac{\epsilon_m}{\epsilon_v + \epsilon_m},
\end{equation}
while $1-y$ reads:
\begin{equation}
1 - y = \frac{\epsilon_v}{\epsilon_v + \epsilon_m}.
\end{equation}
When $y=1$, the situation corresponds to the absence of vacuum energy, while $y=0$ represents a scenario in which the fluid is entirely composed of vacuum energy.

%-------------------------------------------------------------------
\subsection{Constant ratio of ordinary matter and vacuum}
\label{sec:yconst}
%-------------------------------------------------------------------

The TOV equations can be reformulated using only variables associated with ordinary matter and the $y$ fraction. This approach allows us to rewrite Eq. \eqref{mass} as follows:
\begin{equation}
\frac{dm}{dr} = 4 \pi r^2 (\epsilon_m + \epsilon_v) = 4 \pi r^2 \frac{\epsilon_m}{y}.
\end{equation}
When $y$ remains constant, the vacuum energy density is a fixed fraction of $\epsilon_m$, and, as such, $\epsilon_v=\epsilon_v(\epsilon_m)$, and Eq. \eqref{eqh} can be rewritten using the following expressions:
\begin{eqnarray}
\frac{dp}{dr} &=& \frac{dp_m}{dr} - \frac{d \epsilon_v}{dr} = \frac{dp_m}{dr} - \frac{1-y}{y} \frac{d \epsilon_m}{dr},\\
\frac{d \epsilon_m}{dr} & = & \frac{d \epsilon_m}{dp_m} \frac{d p_m}{dr}.
\end{eqnarray}
Substituting these formulas into Eq. \eqref{eqh}, we obtain:
\begin{equation}
\begin{split}
    \frac{dp_m}{dr} = &  -\frac{(p_m + \epsilon_m) \left [m+ 4\pi r^3 \left( p_m - \frac{1-y}{y} \epsilon_m \right)\right]}{r^2 - 2mr} \\
    &\times \frac{1}{\left (1 - \frac{1-y}{y} \frac{d \epsilon_m}{dp_m}\right)}.
    \label{eqhyconst}
\end{split}
\end{equation}
Note that the derivative $\frac{d\epsilon_m}{dp_m}$ is now required in order to  solve the complete set of equations.

The denominator of Eq. \eqref{eqhyconst} must always be positive to ensure a negative pressure gradient within the star. Consequently, we require $y$ to fulfill the condition:
\begin{align}
\left(1 - \frac{1-y}{y} \frac{d\epsilon_m}{dp_m}\right) > 0.
\end{align}
Recalling that the speed of sound in matter is given by $c_s^2 = \frac{dp_m}{d\epsilon_m}$ (in units of the speed of light), we derive a lower limit for $y$:
\begin{align}
y > \frac{1}{c_s^2+1}.
\end{align}
Due to causality, $c_s < 1$, setting an absolute lower limit of $y > 0.5$. However, in practical terms, many realistic EOSs for nuclear matter display a speed of sound below the conformal limit of $1/\sqrt{3}$, suggesting $y > 0.75$. Consequently, in the scenario of constant $y$, a maximum fraction of $25\%$ of vacuum energy would be feasible at each layer of the star.

%-------------------------------------------------------------------
\subsection{Variable fractions of ordinary matter and vacuum}
\label{yvar}
%-------------------------------------------------------------------

We will now shift our focus to the broader scenario where the fraction $y$ of ordinary matter exhibits non-trivial variations throughout the star's interior. In this context, the derivative of the total pressure can be expressed as:
\begin{equation}
\frac{dp}{dr} = \frac{dp_m}{dr} - \frac{d \epsilon_v}{dr} = \frac{dp_m}{dr} - \frac{1-y}{y} \frac{d \epsilon_m}{dr} + \frac{\epsilon_m}{y^2}\frac{dy}{dr}.
\end{equation}
Assuming, for the sake of simplicity, that the fraction $y$ depends solely on the energy density of ordinary matter ($y = y(\epsilon_m)$), we obtain:
\begin{equation}
\frac{dp}{dr} = \frac{dp_m}{dr}  \left[ 1 +  \frac{d\epsilon_m}{dp_m} \left( - \frac{1-y}{y} + \frac{\epsilon_m}{y^2}\frac{dy}{d\epsilon_m} \right)\right].
\end{equation}
By replacing the previous expression in Eq. \eqref{eqh}, one obtains:
\begin{equation}
\begin{split}
\frac{dp_m}{dr}  \left[ 1 +  \frac{d\epsilon_m}{dp_m} \left( - \frac{1-y}{y} + \frac{\epsilon_m}{y^2}\frac{dy}{d\epsilon_m} \right)\right]   =  \qquad \qquad \\  
\qquad =  -\frac{(p_m + \epsilon_m) \left [m+ 4\pi r^3 \left( p_m - \frac{1-y}{y} \epsilon_m \right)\right]}{r^2 - 2mr} .
\end{split}
\label{eq:TOV_variable_y}
\end{equation}

\begin{figure}[tb]
\centering    
\includegraphics[width=0.32\textwidth]{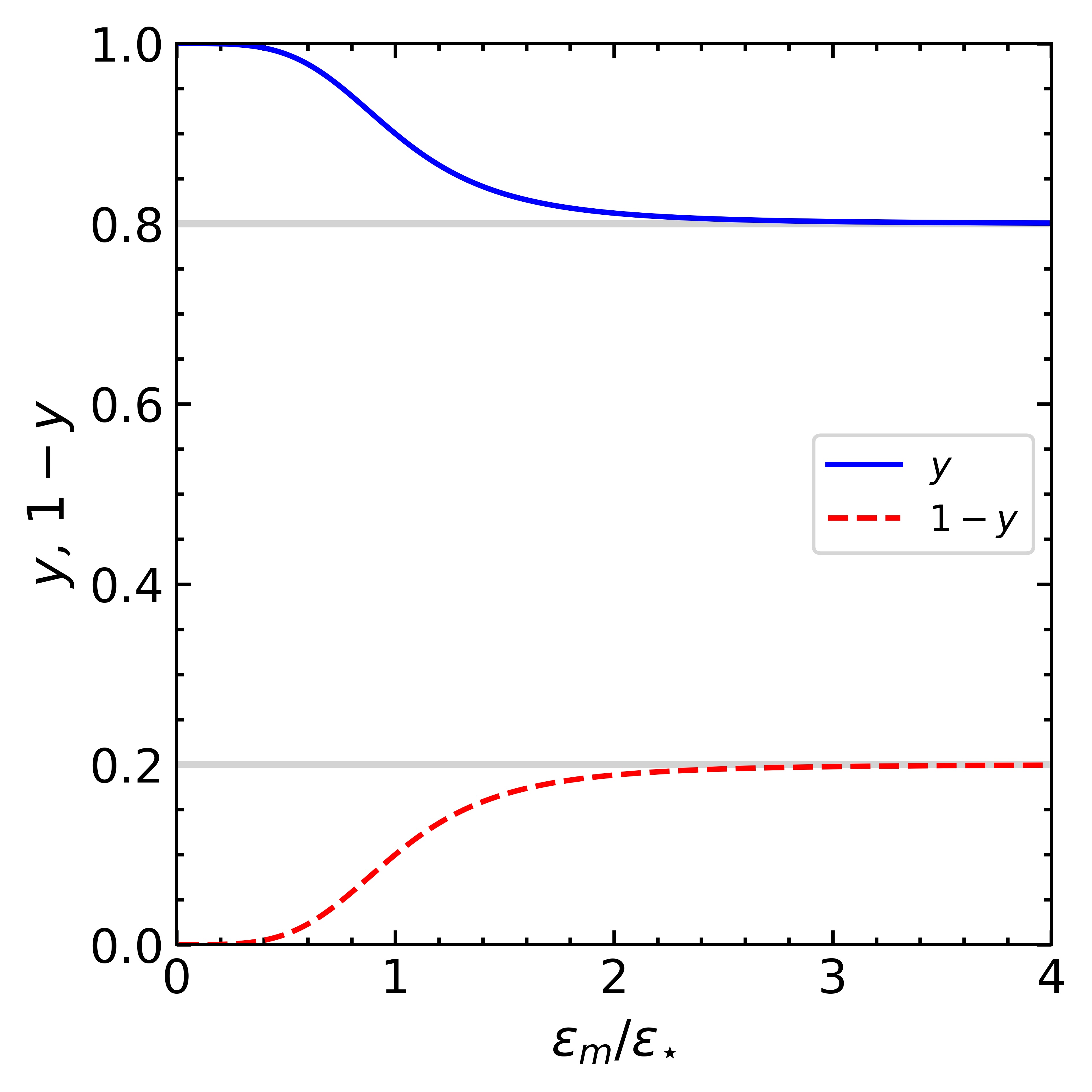}
\caption{Fractions of ordinary matter, $y$, and vacuum energy, $(1-y)$, as functions of $\epsilon_m / \epsilon_{\star}$ for $\beta = 0.8$. Horizontal lines indicate the asymptotic limits, $\beta$ for $y$ and $1-\beta$ for $(1-y)$. At low densities, $y$ tends toward 1 and $(1-y)$ toward zero, evidencing the complete disappearance of vacuum energy.  At high densities, the vacuum energy fraction asymptotically approaches $1-\beta$, representing the maximum fraction allowable under this ansatz.}
\label{yrcomp}
\end{figure}

To describe the variation of $y$ with the energy density of ordinary matter, we propose the following ansatz:
\begin{equation}
y(\epsilon_m) = 1 + \frac{\beta - 1}{1 + \left(\frac{\epsilon_\star}{\epsilon_m}\right)^4},
\label{eq:variable_y}
\end{equation}
where $\beta$ and $\epsilon_\star$ are free parameters. 
The behavior of the function $y(\epsilon_m)$ is illustrated in Fig. \ref{yrcomp}. When $\beta = 1$, the function simplifies to a constant, $y(\epsilon_m) = 1$, representing the absence of dark energy throughout the stellar interior. For $\beta$ values between 0 and 1, the function approaches a limit of 1 as $\epsilon_m$ tends towards zero, indicating that vacuum energy vanishes near the stellar surface. As $\epsilon_m$ approaches infinity, the function tends towards $\beta$, representing a scenario where the fraction of vacuum energy density attains the maximum possible value allowed by the ansatz. The parameter $\epsilon_\star$ acts as a scaling factor that governs the range of $\epsilon_m$ over which $y$ differs significantly from the limiting values $\beta$ and 1. Note that for $\epsilon_\star = 0$, the value of $y$ equals $\beta$ at any energy density. In other words, the ansatz of Eq. \eqref{eq:variable_y} simplifies to the scenario of constant $y$ as described in Sec. \ref{sec:yconst}.

By using Eq. \eqref{eq:variable_y}, we can easily determine the derivative $dy/d\epsilon_m$, which is used in Eq. \eqref{eq:TOV_variable_y}:
\begin{equation}
\frac{dy}{d\epsilon_m} = \frac{4 (\beta-1)(\epsilon_\star/\epsilon_m)^4}{\epsilon_m[1 + (\epsilon_\star/\epsilon_m)^4]^2} .
\end{equation}

% FIGURE 1
%%%%%%%%%%%%%%%%%%%%%%%%%%%%%%%%%%%%%%%%%%%%%%%%
%%%%%%%%%%%%%%%%%%%%%%%%%%%%%%%%%%%%%%%%%%%%%%%%

%-------------------------------------------------------------------
\section{Equations of state for ordinary matter}
\label{sec:EOS}
%-------------------------------------------------------------------

In this work, we employ a piecewise polytropic EOS to effectively represent complex, state-of-the-art nuclear EOSs. This approach simplifies intricate EOSs by approximating them with segmented polytropic relations, offering a versatile method for modeling dense matter within neutron stars. The number of segments and the polytropic indices in each segment can be adjusted to best fit different EOSs, retaining sufficient accuracy and enabling the representation of diverse microphysical approaches in a practical and flexible manner. Here, we adopt the Generalized Piecewise Polytropic (GPP) model as detailed in Ref. \cite{o2020parametrized}, which uses an ansatz that imposes continuity not only in pressure and energy density but also in the speed of sound.

The pressure and the energy density of each segment are written as functions of the rest mass density $\rho$ as follows:
\begin{align}
p_m(\epsilon_m)     &= K \rho_m^{\Gamma} + A,\\
\epsilon_m(\rho_m)  &= \frac{K}{\Gamma - 1} \rho_m^{\Gamma} + (1 + a) \rho_m - A,
\end{align}
from which we derive:
\begin{align}
\frac{d\epsilon_m}{dp_m} = \frac{1}{\Gamma - 1} + \frac{(1+a)}{\Gamma K \rho_m^{\Gamma - 1}}.
\end{align}

The coefficients of adjacent segments are related by imposing continuity and differentiability of $p$ and $\epsilon$ at the dividing density $\rho_i$: 
\begin{align}
K_{i+1} &= K_i \frac{\Gamma_i}{\Gamma_{i+1}} \rho_i^{\Gamma_i - \Gamma_{i+i}}, \\
A_{i+1} &= A_i + \left(1- \frac{\Gamma_i}{\Gamma_{i+1}} \right) K_i \rho^{\Gamma_i}_i, \\
a_{i+1} & =a_i+\Gamma_i \frac{\Gamma_{i+1}-\Gamma_i}{\left(\Gamma_{i+1}-1\right)\left(\Gamma_i-1\right)} K_i \rho_i^{\Gamma_i-1}.
\end{align}

From the extensive set of EOSs reproduced in Ref. \cite{o2020parametrized}, we have selected only those that match the predictions of chiral effective field theory (EFT) interactions. Chiral EFT provides a framework for a systematic expansion of nuclear forces at low momenta, explaining the hierarchy of two-, three-, and weaker higher-body forces \cite{Coraggio:2012ca, Gandolfi:2011xu, Holt:2012yv, Hebeler:2009iv, Sammarruca:2012vb, Tews:2012fj}. Currently, microscopic calculations based on chiral EFT interactions allow for a reliable determination of the properties of neutron star matter up to densities somewhat above the nuclear saturation density $n_0$. Specifically, the pressure is currently known to be roughly within $\pm 20\%$ accuracy at $n_0$ \cite{Tews:2012fj, Hebeler:2009iv}.

%%%%%%%%%%%%%%%%%%%%%%%%%%%%%%%%%%%%%%%%%%%%%%
\begin{figure}[tb]
\centering
\includegraphics[width=0.43\textwidth]{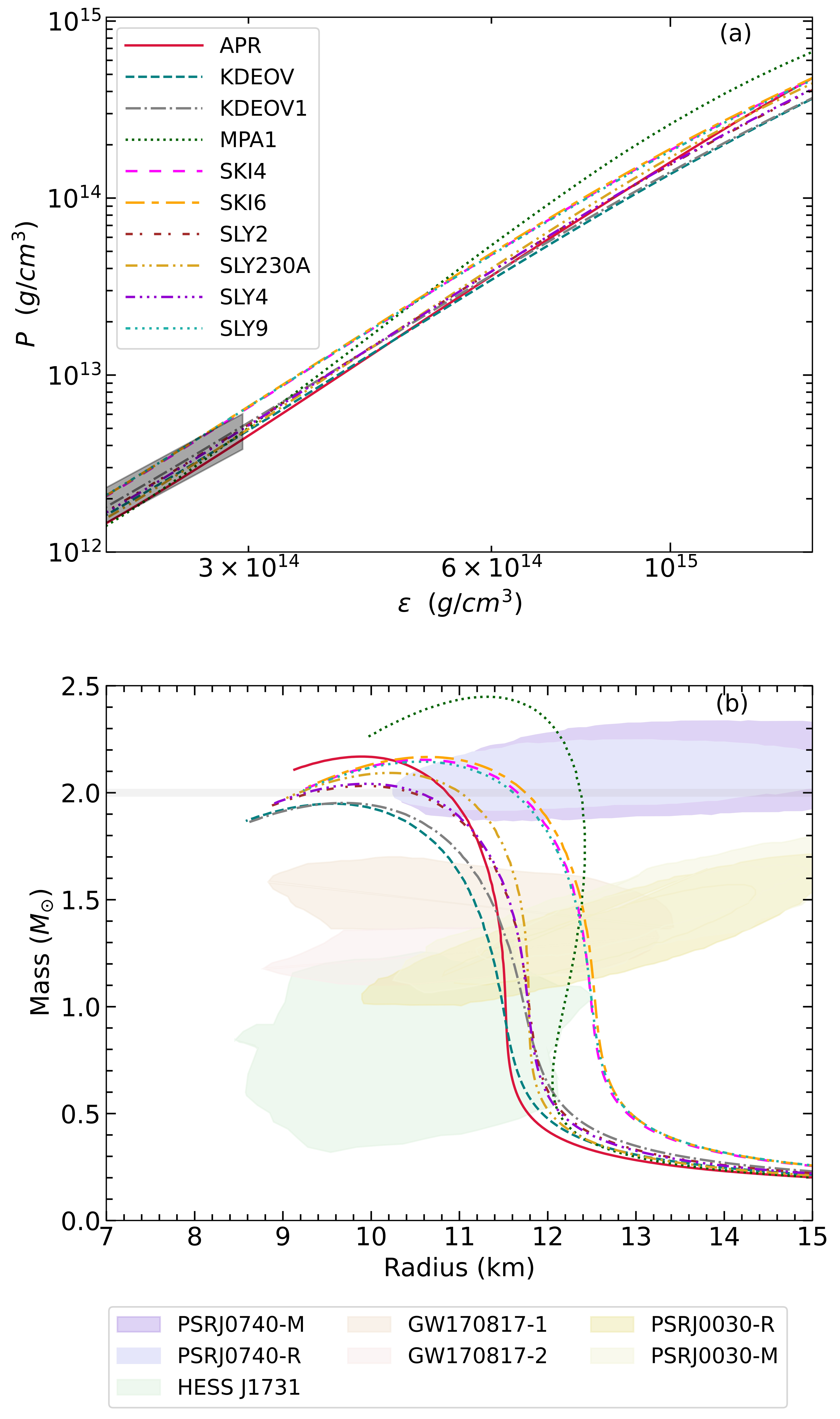}
\caption{(a) Selected EOSs for ordinary matter used in this work. The gray shaded area in the lower-left corner represents the constraint imposed by chiral effective field theory up to $1.1\,n_0$ \citep{Hebeler:2013nza, Annala:2020efq}, a constraint satisfied by all of our hadronic EOSs. (b) Mass-radius relationship for the selected EOSs (without any vacuum contribution), illustrating the differences in predicted neutron star sizes and masses. Overlaying the theoretical curves are shaded regions,  corresponding to observational constraints from $\sim 2 M_{\odot}$ pulsars,  NICER observations, the GW170817 event detected by LIGO/Virgo, and the very low mass object HESS J1731.}
\label{fig:eos_no_vacuum}
\end{figure}
%%%%%%%%%%%%%%%%%%%%%%%%%%%%%%%%%%%%%%%%%%%%%%

The selected models are presented in Fig. \ref{fig:eos_no_vacuum}(a), along with the region delineating the constraints imposed by chiral EFT. Each model reflects different assumptions regarding the nature of the forces between nucleons and the potential emergence of exotic particles such as hyperons at extreme densities. The APR EOS is based on variational calculations using potentials resulting from fits to nucleon-nucleon scattering and properties of light nuclei \cite{Akmal:1998cf}. The KDE0V, KDE0V1, SKI4, SKI6, SLY2, SLY230A, SLY4, and SLY9 models are part of the Skyrme interaction family \cite{Dutra:2012mb}. In Skyrme-type models, the effective mass and nuclear potential are parametrized to reproduce properties of bulk nuclear matter and/or finite nuclei. Each variant of the Skyrme interaction represents a different parameter set, fitted to specific nuclear properties and experimental data. The MPA1 EOS is based on relativistic Brueckner-Hartree-Fock calculations, with a particular focus on contributions resulting from tensor interactions due to the exchange of $\pi$ and $\rho$ mesons and the dependence on neutron-proton asymmetry \cite{Muther:1987xaa}.

In Fig. \ref{fig:eos_no_vacuum}(b), the mass-radius relationship for neutron stars is shown using the EOSs presented in Fig. \ref{fig:eos_no_vacuum}(a), without considering any vacuum energy effects.  The KDE0V and KDE0V1 models fail to meet the observational constraints imposed by  PSR J0740+6620 and $2 M_{\odot}$ pulsar measurements. However, the remaining EOSs generally comply with all astrophysical constraints, despite some satisfying the HESS J1731 constraint only marginally. Subsequent figures will particularly emphasize the APR and MPA1 models.  The APR model is somewhat less stiff than the MPA1 model, resulting in a lower predicted maximum mass. Furthermore, stellar radii within the APR model are considerably smaller than those predicted by the MPA1 model.  This choice will provide deeper insight into the role of vacuum energy in EOSs with different degrees of stiffness.

%%%%%%%%%%%%%%%%%%%%%%%%%%%%%%%%%%%%%%%%%%%%%%
\begin{figure*}[tb]
\centering
\includegraphics[width=1.\textwidth]{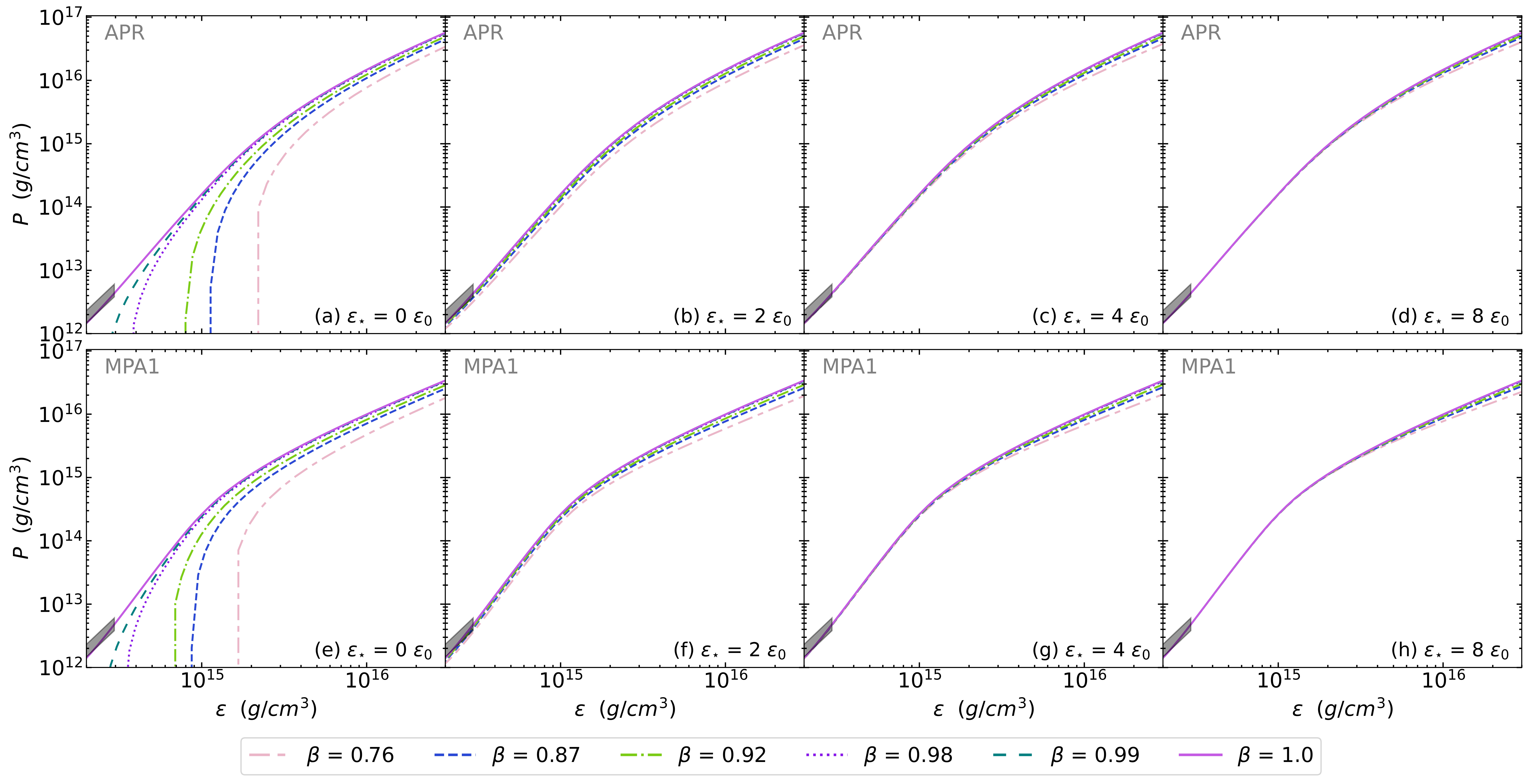}
\caption{Effect of vacuum energy on APR and MPA1 EOSs. The dashed and dotted curves incorporate the effect of vacuum energy for different $\beta$ (or $y$ in the constant case). Note that there are significant changes in the stiffness of the EOS, especially at low-density regimes for $\epsilon_\star = 0 \epsilon_0$ (constant $y$). This effect does not contradict nuclear physics, as the modified EOS only becomes relevant in strong gravitational fields. In general, the EOS exhibits only slight softening at high densities.} 
\label{fig:eos_with_vacuum}
\end{figure*}
%%%%%%%%%%%%%%%%%%%%%%%%%%%%%%%%%%%%%%%%%%%%%%

%--------------------------------------
\section{Results}
\label{sec:results}
%--------------------------------------

In this section, we will solve the stellar structure equations described in the previous sections, using the set of EOSs outlined in Sec. \ref{sec:EOS}. This will involve adopting different proportions and distributions of vacuum energy within the interior of the star, as explored in Sec. \ref{sec:results_eos}. In Sec. \ref{sec:results_MR}, our focus will be on the mass-radius relationship, while in Sec. \ref{sec:results_tidal}, we will analyze tidal deformabilities.

%-------------------------------------------------------------------
\subsection{Vacuum energy effects in the equation of state}
\label{sec:results_eos}
%-------------------------------------------------------------------

Let us analyze in detail how two of the previously selected nuclear matter EOSs, namely APR and MPA1, are altered when vacuum energy is incorporated according to the prescriptions discussed earlier.
As pointed out at the end of Sec. \ref{yvar},  the scenario with a constant $y$ corresponds to the case of variable $y$ with $\epsilon_\star = 0$, and this notation will henceforth denote the constant $y$ model. This scenario is depicted in Fig. \ref{fig:eos_with_vacuum}(a) for the APR EOS and in Fig. \ref{fig:eos_with_vacuum}(e) for the MPA1 EOS. The remaining panels of Fig. \ref{fig:eos_with_vacuum} show the impact of vacuum energy within the variable $y$ scenario, employing different values for $\epsilon_\star$, specifically $2\epsilon_0$, $4\epsilon_0$, and $8\epsilon_0$, with $\epsilon_0 = 2.7 \times 10^{14} \mathrm{~g/cm^3}$ representing the nuclear saturation density. These values were chosen to set the onset of the vacuum energy contribution at low, intermediate, or high densities, respectively.

As expected, in all the cases presented in Fig. \ref{fig:eos_with_vacuum}, the EOS softens upon the addition of vacuum energy due to its negative pressure contribution. It is also observed that the smaller the parameter $\beta$, the more pronounced the softening of the EOS. This trend suggests that increasing the contribution of vacuum energy will progressively decrease the maximum mass and shift the mass-radius curves toward smaller radii, as will indeed be demonstrated in Sec. \ref{sec:results_MR}. An interesting aspect of the models with constant $y$ is that, as shown in Figs. \ref{fig:eos_with_vacuum}(a) and \ref{fig:eos_with_vacuum}(e), the total pressure vanishes at quite high densities, and is negative below $10^{14}-10^{15} \mathrm{g/cm^3}$. This occurs because, in the constant $y$ scenario, the vacuum contribution to the pressure, represented by $\epsilon_v = (1-y) y^{-1} \epsilon_m$, remains substantial even at low pressures and eventually surpasses the matter pressure at a relatively high density. Conversely, as shown in Fig. \ref{fig:eos_with_vacuum}, in the variable $y$ scenario, the effect of vacuum energy is mitigated at densities lower than approximately $\epsilon_\star$ (also illustrated in Fig. \ref{yrcomp}), preventing the total pressure from becoming negative in most cases.

%-------------------------------------------------------------------
\subsection{The influence of vacuum energy on stellar structure}
\label{sec:results_MR}
%-------------------------------------------------------------------

%%%%%%%%%%%%%%%%%%%%%%%%%%%%%%%%%%%%%%%%%%%%%%
\begin{figure*}[tb]
\centering     
\includegraphics[width=1.\textwidth]{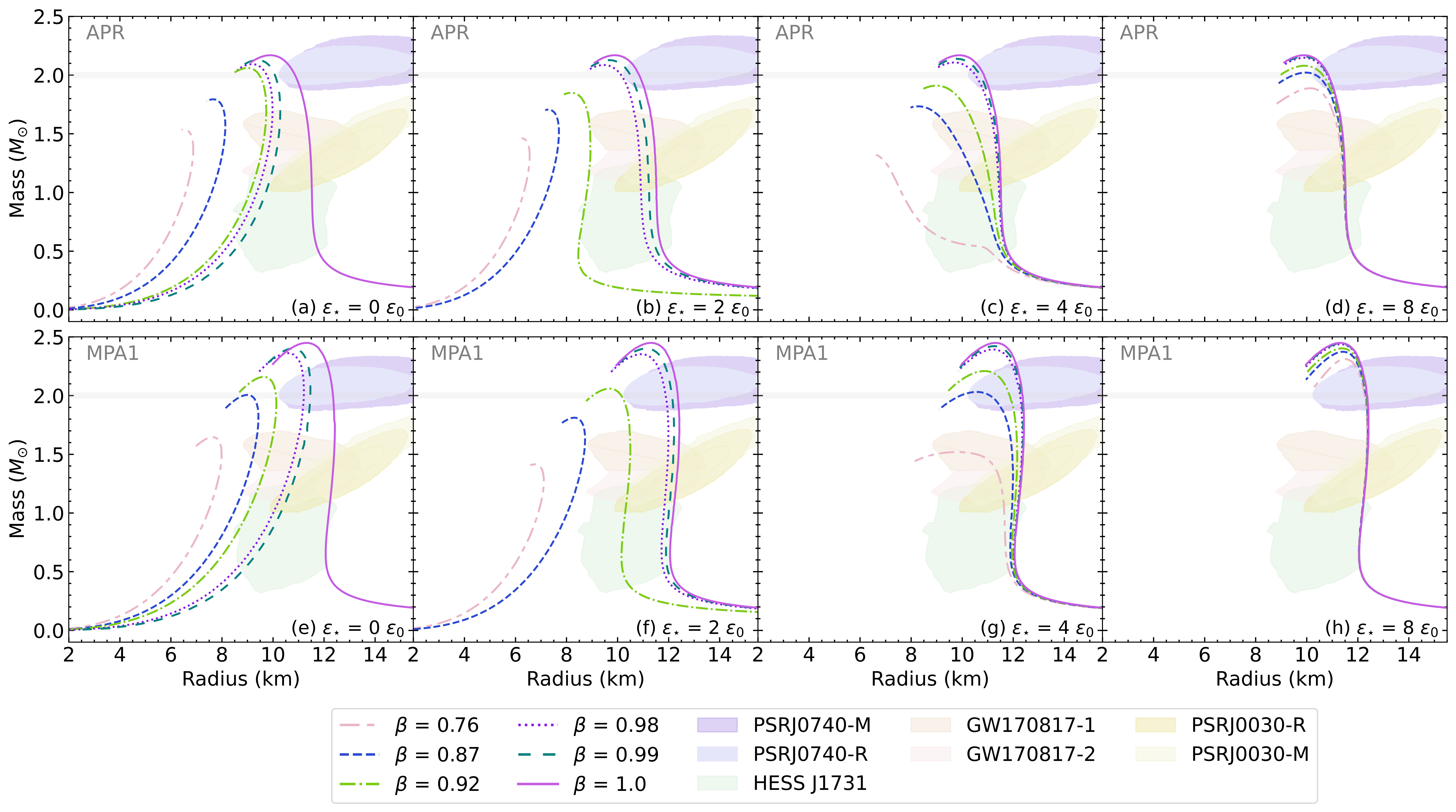}
\caption{Mass-Radius relation for APR and MPA1 with the inclusion of vacuum energy related with the ordinary matter in the star by a function $y$, and observational constraints (GW170817 from \cite{raithel2018tidal}, GW190425 from \cite{abbott2020gw190425}, PSR J0030+0451 from \cite{miller2019psr}, PSR J0348+0432 from \cite{antoniadis2013massive}, PSR J1614+2230 from \cite{demorest2010two} and PSR J0740+6620 from \cite{cromartie2020relativistic}).}
\label{fig:M_vs_R}
\end{figure*}
%%%%%%%%%%%%%%%%%%%%%%%%%%%%%%%%%%%%%%%%%%%%%%

We have calculated the mass-radius relationship for the set of nuclear matter EOSs selected previously. For brevity and focus, we will only show plots for two of these EOSs, specifically APR (Figs. \ref{fig:M_vs_R}(a)--\ref{fig:M_vs_R}(d)) and MPA1 (Figs. \ref{fig:M_vs_R}(e)--\ref{fig:M_vs_R}(h)).
Each panel corresponds to a different scenario regarding the proportion of vacuum energy. Observational constraints are overlaid on the curves, shown by the shaded regions and labeled with the names of the observed objects. Figs. \ref{fig:M_vs_R}(a) and \ref{fig:M_vs_R}(e) illustrate the mass-radius relationship for neutron stars under the assumption of a constant $y$ ($\epsilon_{\star} = 0$). Figs. \ref{fig:M_vs_R}(b)--\ref{fig:M_vs_R}(d)  and \ref{fig:M_vs_R}(f)--\ref{fig:M_vs_R}(h) display the results with density-dependent values of $y$ under varying thresholds of $\epsilon_{\star}$, specifically at $2\epsilon_0$, $4\epsilon_0$, and $8\epsilon_0$, respectively. Within each panel various selections for the parameter $\beta$ are explored.

%######################################################

Across all examined scenarios, a consistent trend emerges:  both radii and masses tend to be smaller as the amount of vacuum energy is increased. This effect is particularly pronounced in the scenario where the parameter $y$ is held constant  (Figs. \ref{fig:M_vs_R}(a) and \ref{fig:M_vs_R}(e)) because vacuum energy exerts a substantial influence on the EOS throughout the entire density spectrum.  On the other hand, in the scenario with density dependent $y$  the effect is substantially reduced as the parameter $\epsilon_{\star}$ becomes larger, simply because the presence of vacuum energy is restricted to higher density regimes, thus exerting a lesser impact on the star's structure (cf. Fig. \ref{yrcomp}). For instance, with the APR EOS, when $\epsilon_{\star} = 4 \epsilon_0$, the results for $\beta \lesssim 0.96$ become incompatible with the  $2M_{\odot}$ constraint. However, for $\epsilon_{\star} = 8 \epsilon_0$, a significantly broader range of $\beta$ values ($1 < \beta \lesssim 0.87$) yield $M-R$ relationships that are consistent with the  $2M_{\odot}$ constraint. 

Another significant effect of vacuum energy is related to the qualitative shape of the mass-radius curves. In the case of constant $y$, even a small fraction of vacuum energy leads to a qualitative change in the curves, making them resemble those typical of self-bound stars, such as strange quark stars \cite{Alcock:1986hz, Lugones:2002zd, VasquezFlores:2017uor, Lugones:2023zfd}. In fact, unlike standard neutron stars, which exhibit larger radii at lower masses, strange quark stars show a decrease in radius with decreasing mass. This effect is observed in Figs. \ref{fig:M_vs_R}(a) and \ref{fig:M_vs_R}(e) with vacuum energy. For density-dependent $y$, the sensitivity to vacuum energy strongly depends on the value of $\epsilon_\star$. When vacuum appears at high densities, its impact is similar to what is observed in hybrid stars, which feature a quark matter core surrounded by nuclear matter \cite{Lenzi:2012xz, Mariani:2023kdu}: There is a decrease in $M_{\mathrm{max}}$ and a shift of the $M-R$ curves towards smaller radii, particularly noticeable in Figs. \ref{fig:M_vs_R}(c), \ref{fig:M_vs_R}(d), \ref{fig:M_vs_R}(g), and \ref{fig:M_vs_R}(h). This shift affects objects with higher masses and leaves low mass neutron stars unaffected, ensuring the qualitative shape of the curve remains unchanged. On the other hand, when sufficiently large values of $\beta$ along with sufficiently small values of $\epsilon_\star$ are adopted, it is still possible to achieve curves with the same shape as self-bound stars (see Figs. \ref{fig:M_vs_R}(b), \ref{fig:M_vs_R}(f) with $\beta = 0.76, 0.87$). However, in all the cases we analyzed, these curves are incompatible with the $2M_{\odot}$ constraint.

The behavior described above can be understood as follows: The vacuum energy contributes additively to density but negatively to pressure: $\epsilon = \epsilon_m + \epsilon_v$ and $p = p_m - \epsilon_v$. Therefore, vacuum energy invariably softens the total EOS. When vacuum energy is restricted to the stellar core, this softening leads to the well-known effects of reducing $M_{\mathrm{max}}$ and shifting the mass-radius curves towards smaller radii \cite{Lenzi:2012xz, Lugones:2021bkm}. However, as the influence of vacuum energy extends to lower densities, it begins to significantly bind the normal matter at the star’s surface. Specifically, the condition $p(R) = 0$ (which implies $p_m(R) = \epsilon_v(R)$) allows for configurations where the matter pressure at the surface is non-vanishing, while the total pressure remains zero. This effect ``truncates'' the star compared to models without vacuum energy, resulting in a much smaller radius. This mechanism is similar to that which shapes strange quark stars, the only difference being that the vacuum energy in strange quark stars originates from strong interactions \cite{Lugones:2002zd, Lugones:2002vd, VasquezFlores:2017uor}.

%######################################################

Observed masses and radii of neutron stars can provide valuable constraints on the amount of vacuum energy they contain.  These constraints, however, depend on the specific EOS employed. EOSs that are close to exceeding the limits set by astrophysical observations will become incompatible with these observations, even with minor additions of vacuum energy to the matter EOS.
In the case of constant $y$, radii are strongly reduced even for tiny additions of vacuum energy. As a consequence, the radius limits set by NICER are the first observational restrictions to be violated by models with vacuum energy. For example, with {$y = 0.995$}, the APR curves are pushed beyond the accepted radius limits of PSR J0740 and PSR J0030. For the MPA1 EOS, the impact is similarly substantial, but because the curve without vacuum energy is positioned much further to the right, reflecting its stiffer nature, the $\beta$ value at which it fails to meet NICER's constraints is lower: {$y \approx 0.92$}. 
In the case of density-dependent $y$, the decrease in radius tends to be less pronounced, particularly in models where vacuum energy is confined to the stellar core. Consequently, the first astrophysical constraint to be infringed by models with vacuum energy can vary; sometimes it is the NICER constraint and other times it is the two solar mass limit.

All the features previously described for APR and MPA1 are also corroborated by the results we obtained for the other EOSs included in our analysis. To summarize the findings across these EOSs, we will focus primarily on the $2M_{\odot}$ constraint, as it is the most robust among the astrophysical restrictions. In Table \ref{table:y_limits}, we present the minimum $\beta$ required for compliance with the $2M_{\odot}$ constraint for all EOSs and for all selected vacuum energy density thresholds $\epsilon_{\star}$. As anticipated, EOSs that are already near the $2M_{\odot}$ maximum mass threshold in the absence of vacuum energy present a higher $\beta_{\mathrm{min}}$.

\begin{table}[tb]
\renewcommand{\arraystretch}{1.15} 
\centering
\begin{tabular}{|cc||cccc|}
\hline
EOS & $M^0_\mathrm{max}$ &  \multicolumn{4}{c|}{$\beta_{\mathrm{min}}$} \\
\cline{3-6}
& & $\epsilon_{\star} =0$ & $\epsilon_{\star} = 2 \epsilon_0$ & $\epsilon_{\star} = 4 \epsilon_0$ & $\epsilon_{\star} = 8 \epsilon_0$ \\
\hline
APR  & 2.170 %2.057 
& 0.970 & 0.970 & 0.965 & 0.870 \\
MPA1 &2.465& 0.870 & 0.910 & 0.870 & 0.760$^*$ \\
SKI4 &2.170& 0.970 & 0.970 & 0.960 & 0.870\\
SKI6 &2.191& 0.970 & 0.970 & 0.950 & 0.860 \\
SLY2 &2.055& 0.990 & 0.990 & 0.985 & 0.980 \\
SLY230A &2.100& 0.980 & 0.980 & 0.975 & 0.930 \\
SLY4 &2.053& 0.990 & 0.990 & 0.985 & 0.980 \\
SLY9 &2.157& 0.980 & 0.980 & 0.970 & 0.870 \\
\hline
\end{tabular}
\caption{For each EOS, this table presents the minimum values of the parameter $\beta$ required for agreement with the  $2 M_{\odot}$  constraint across different vacuum energy density thresholds: $\epsilon_{\star} = 0, 2 \epsilon_0, 4 \epsilon_0, 8 \epsilon_0$.  The column $M^0_\mathrm{max}$ indicates the maximum mass predicted by each EOS in the absence of vacuum energy.}
\label{table:y_limits}
\end{table}

%-------------------------------------------------------------------
\subsection{Tidal deformability}
\label{sec:results_tidal}
%-------------------------------------------------------------------

%%%%%%%%%%%%%%%%%%%%%%%%%%%%%%%%%%%%%%%%%%%%%%
\begin{figure*}[ht!]
\centering     %%% not center
\includegraphics[width=1.\textwidth]{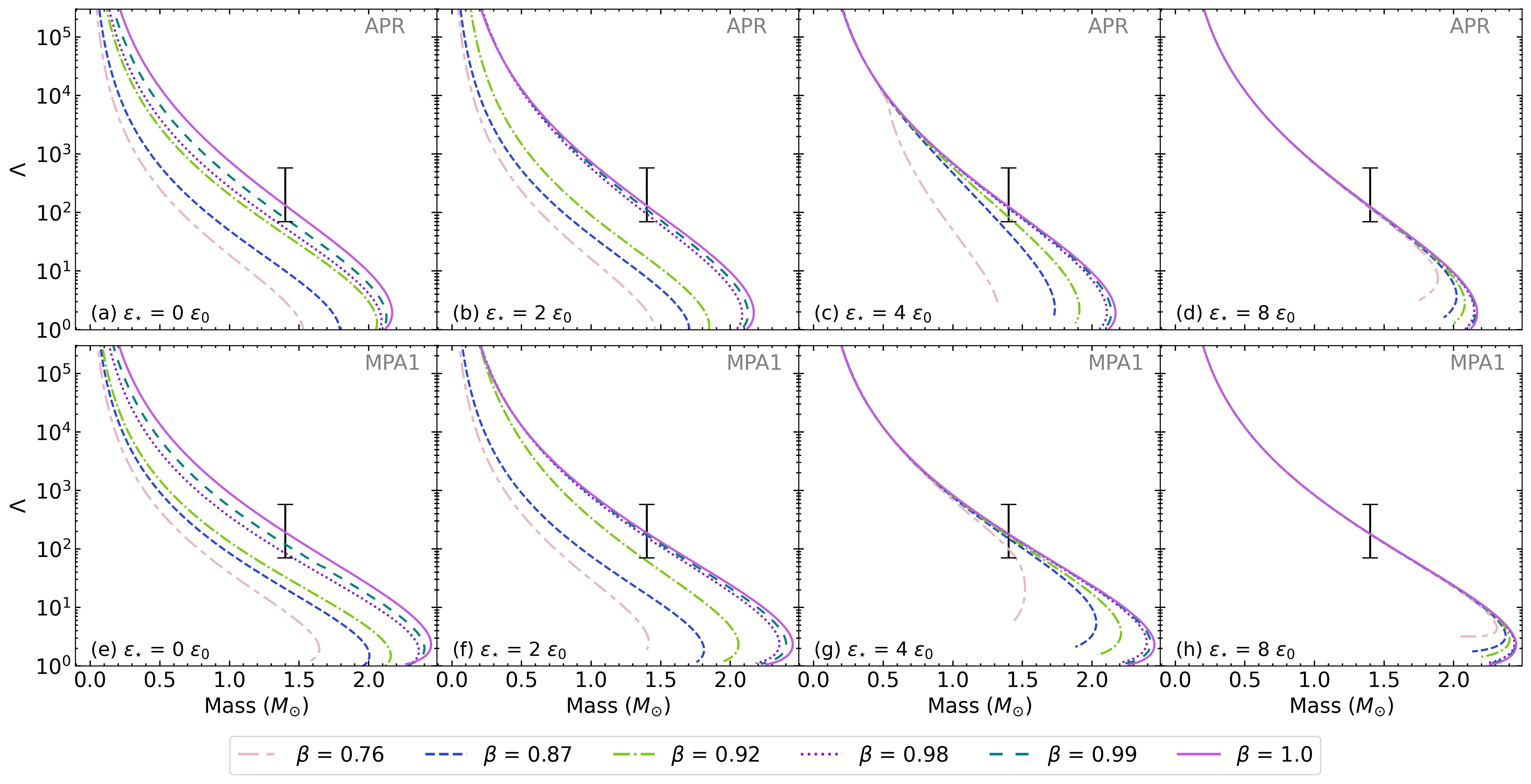}
\caption{Tidal deformability for the  APR and MPA1 EOSs with vacuum energy. The color coding and organization are the same as in Figs. \ref{fig:eos_with_vacuum} and \ref{fig:M_vs_R}. The constraint on $\Lambda_{1.4}$ from the merger event GW170817 \cite{Abbott:2018exr} is shown as vertical error bars in each panel. }
\label{tidal}
\end{figure*}
%%%%%%%%%%%%%%%%%%%%%%%%%%%%%%%%%%%%%%%%%%%%%%

Neutron stars experience significant tidal deformation during the inspiral phase of a neutron star-neutron star merger leaving a detectable imprint on the observed gravitational waveform of the merger.  The extent of this effect is quantifiable in terms of the star's dimensionless tidal deformability,  which characterizes the induced mass quadrupole moment in response to an external tidal field.  Specifically, the induced mass-quadrupole moment $Q_{ij}$ and the applied tidal field $\epsilon_{ij}$ are related to this parameter by $Q_{ij} = -\Lambda M^5 \epsilon_{ij}$. 

The dimensionless tidal deformability parameter can be expressed as:
\begin{equation}
\Lambda = \frac{2}{3} k_2 C^{-5},
\label{eq:tidal_deformability}
\end{equation}
where $C \equiv M/R$ is the compactness of the star and $k_2$ is the tidal Love number, which, in turn, is given by 
\begin{equation}
\begin{aligned}
k_{2} &=   \frac{8C^{5}}{5} \left(1-2C\right)^{2} \left[2 + 2C \left(y_R-1\right)-y_R\right]  \\
& ~   \times \bigl[ 4C^{3}\left[13-11y_R+C\left(3y_R-2\right)+2C^{2} \left(1+y_R\right)\right] \\
& ~  +3\left(1-2C\right)^{2}\left[2-y_R+2C \left(y_R-1\right)\right] \log\left(1-2C\right) \\
& ~ +  2C \left[6-3y_R+3C \left(5y_R-8\right)\right]  \bigr]^{-1}. 
\label{eq:love}
\end{aligned}
\end{equation}
Here, $y_R\equiv y(R)$, being $y(r)$ the solution of the following first-order differential equation  \cite{Postnikov:2010yn}:   
\begin{equation}
r \frac{dy(r)}{dr}+y(r)^2+y(r)F(r)+r^2Q(r)=0 .
\label{eq:tidal}
\end{equation}
In this equation, the coefficients are: 
\begin{eqnarray}
F(r) &=& \left[1-4\pi r^2(\epsilon-p)\right]\left[1-\frac{2m}{r}\right]^{-1}, \\
\label{functionQ}
Q(r) &=& 4\pi\left[5\epsilon +9p +\frac{\epsilon+p}{c_s^2}-\frac{6}{4\pi r^2}\right]\left[1-\frac{2m}{r}\right]^{-1} \nonumber\\ 
&& -\frac{4m^2}{r^4}\left[1+\frac{4\pi r^3 p}{m}\right]^2\left[1-\frac{2m}{r}\right]^{-2}, 
\end{eqnarray}
where $c_s^2$ is the squared speed of sound, which within the present model can be expressed as:
\begin{align}
c_s^2 \equiv \frac{dp}{d\epsilon} = \frac{\left(\frac{\epsilon_m}{y^2} \frac{dy}{d\epsilon_m} - \frac{1}{y} + 1 + \frac{dp_m}{d\epsilon_m}\right)}{\left(-\frac{\epsilon_m}{y^2} \frac{dy}{d\epsilon_m} + \frac{1}{y}  \right)} .
\end{align}
As in previous sections, $m = m_m + m_v$, $p = p_m + p_v$, and $\epsilon = \epsilon_m + \epsilon_v$. The boundary condition for Eq.~\eqref{eq:tidal} at $r=0$ is given by  $y(0)=2$. 
In summary, the tidal Love number can be obtained once an EOS is supplied and the TOV equations together with Eq. \eqref{eq:tidal} are integrated.

The primary constraint on the value of $\Lambda$ currently stems from the detection of gravitational waves from the binary neutron star merger event GW170817 by LIGO/Virgo \cite{LIGOScientific:2017vwq, Abbott:2018exr, LIGOScientific:2018hze}. Assuming that both objects have spins within the range observed in Galactic binary neutron stars and that they share the same EOS, the tidal deformability of a  $1.4 M_{\odot}$  neutron star can be estimated to be $\Lambda_{1.4}=$ $190_{-120}^{+390}$ at the $90 \%$ level \cite{Abbott:2018exr}.   

In Fig.~\ref{tidal} we show $\Lambda$ as a function of $M$ for the EOSs of Fig.~\ref{fig:eos_with_vacuum}. The results indicate that the highest values of tidal deformability for a given mass always correspond to the curves with $\beta = 1$ (no vacuum energy effect). As the amount of vacuum energy increases (indicated by lower $\beta$ values), tidal deformability decreases for a given mass. Also, as the threshold for the onset of vacuum energy effects increases ($\epsilon_*$ goes from 0 to $8 \epsilon_0$), the impact of vacuum energy on tidal deformability becomes less pronounced for both EOS, with the curves beginning to converge, especially for low $M$. 
For the APR EOS, the effect of vacuum energy is more noticeable at lower $\epsilon_*$ thresholds, as evidenced by the significant spread of the $\Lambda$ curves seen in Figs.~\ref{tidal}(a) through \ref{tidal}(c). For the MPA1 EOS, a similar pattern is observed; however, the influence of vacuum energy appears to be less severe compared to the APR EOS. At higher vacuum energy density thresholds, as seen in Figs.~\ref{tidal}(d) and \ref{tidal}(h), the curves for both EOSs behave more similarly. Overall, our results show a clear trend of decreasing tidal deformability with increasing contributions of vacuum energy, particularly at lower energy density thresholds.

We now examine how these calculated values of $\Lambda$ compare with the restriction imposed by the GW170817 merger event. For both EOS, models without vacuum energy ($\beta = 1$) are already consistent with the error bars of GW170817. As the amount of vacuum energy increases, especially in cases with lower $\epsilon_*$ thresholds, the curves diverge substantially from the error bars. For the APR EOS, the broad spread of theoretical curves at lower $\epsilon_*$ thresholds means that only curves with higher $\beta$ values, closer to 1, are consistent with the observational constraints from GW170817. Stiffer models such as MPA1 might accommodate a broader range of vacuum energy contributions while still being consistent with GW170817. At higher vacuum energy density thresholds, with $\epsilon_*$ increasing to $8 \epsilon_0$, the theoretical curves for different $\beta$ tend to coincide  (see Figs.~\ref{tidal}(d) and ~\ref{tidal}(h)). This convergence results in all models becoming compatible with the $\Lambda_{1.4}$ constraints from GW170817, since the variations in the curves caused by the addition of vacuum energy occur at masses significantly larger than $1.4 M_{\odot}$.

%----------------------------------------------
\section{Summary and Conclusions}
\label{sec:conclusions}
%----------------------------------------------

In this work, we explored the role of vacuum energy within neutron stars, assuming it is admixed with ordinary matter. To describe ordinary matter we employed a polytropic representation of a set of state-of-the-art nuclear EOSs, while for vacuum energy, we adopted the standard relation $p_{v}=-\epsilon_{v}$.   To complete the system of equations, we assumed that, at each point of the star, ordinary matter constitutes a fraction $y$ of the total energy density, and vacuum energy constitutes a fraction $1-y$.   The fraction $y$ was specified according to different models. Initially, we assumed that $y$ remains constant throughout the stellar interior. We then adopted a density-dependent approach as defined in Eq.~\eqref{eq:variable_y}, where at low densities $y \rightarrow 1$, indicating the complete disappearance of vacuum energy, and at high densities, $y \rightarrow \beta$, where $\beta$ represents the maximum possible fraction of vacuum energy present in the system.

Our selection of ordinary matter EOSs was specifically aimed to align with the predictions of chiral EFT interactions near the nuclear saturation density. Detailed analysis in Sec. \ref{sec:results_eos} shows that incorporating vacuum energy modifies these EOSs, typically resulting in their softening due to the negative pressure contribution of vacuum energy. The degree of softening intensifies with higher values of the parameter $\beta$. In models maintaining a constant fraction $y$, total pressure vanishes at relatively high densities, as the substantial contribution from vacuum energy surpasses the matter pressure. In contrast, for models with a density-dependent $y$, where vacuum energy's influence disappears at densities below approximately $\epsilon_\star$, the total pressure remains positive in all relevant cases.

We subsequently examined the mass-radius relationship for a set of nuclear matter EOSs, with a particular focus on the APR EOS and the significantly stiffer MPA1 EOS. Across all scenarios, we observed reductions in both stellar radii and masses as the contributions of vacuum energy increased. In models with a constant $y$, even minimal increases in vacuum energy profoundly affect the qualitative shape of the mass-radius curves, rendering them similar to those observed in strange quark stars, where $R \rightarrow 0$ as $M \rightarrow 0$. In contrast, in models with a density-dependent $y$, the changes are more quantitative in nature. Indeed, the impact of vacuum energy mirrors that seen in quark-hadron hybrid stars, i.e. it reduces the maximum masses and shifts the mass-radius curves towards smaller radii. These changes predominantly affect more massive neutron stars while leaving those of lower mass unaffected, thereby preserving the overall shape of the curves.

We evaluated our theoretical curves against observed masses and radii of neutron stars. For models with constant $y$, even minimal increases in vacuum energy significantly reduce radii, leading these models to first infringe the radius limits set by NICER. In models with density-dependent $y$, the radius reduction is generally milder, especially when vacuum energy is concentrated in the stellar core. Therefore, the first astrophysical constraints to be challenged by these models vary, sometimes being the NICER constraints and at other times the two solar mass limit. Given the robustness of the $2M_{\odot}$ constraint, we detailed in Table \ref{table:y_limits} the minimum $\beta$ values needed to satisfy this limit for all selected EOSs across various vacuum energy density thresholds $\epsilon_{\star}$. Predictably, EOSs close to the $2M_{\odot}$ maximum mass threshold without vacuum energy require a higher $\beta_\mathrm{min}$.

Finally, we showed that, at fixed stellar mass, the tidal deformability $\Lambda$ diminishes as the amount of vacuum energy increases. Moreover, as the threshold for vacuum energy effects rises (from $0$ to $8 \epsilon_0$), the influence of vacuum energy on tidal deformability becomes less significant, leading to a convergence of curves, particularly at lower masses. We then juxtaposed the theoretical values of $\Lambda$ against the constraint from the GW170817 event. Our EOSs without vacuum energy match the observational error bars from GW170817. However, as the vacuum energy fraction is increased, particularly at lower $\epsilon_*$ thresholds, the curves significantly deviate from these error bars. At higher $\epsilon_*$, the theoretical curves for different $\beta$ values tend to coincide, especially at low masses, and all of them align with the $\Lambda_{1.4}$ constraint from GW170817.

To conclude, we mention some challenges of our approach and the broader implications of our findings. Notably, the levels of vacuum energy we investigated exceed, by more than 40 orders of magnitude, those derived from cosmological estimates, where vacuum energy density approximates $\epsilon \sim 10^{-29} \mathrm{g/cm}^3$. A significant theoretical challenge lies in elucidating the mechanisms by which dark energy could be significantly concentrated within neutron stars via gravitational interactions. Furthermore, even assuming such a substantial accumulation of dark energy is possible, determining its presence within neutron stars observationally remains problematic. The influence of vacuum energy could be indistinguishable from effects attributed to other complex stellar phenomena, such as those associated with strange quark stars or hybrid stars, both of which are well-documented in existing astrophysical literature. This overlap complicates the interpretation of observational data and highlights the need for innovative approaches to distinguish between these various influences on neutron star behavior.

\section*{Acknowledgments}
LFA was supported by Coordenação de Aperfeiçoamento de Pessoal de Nível Superior (CAPES), Brazil. JASL is partially supported by the Brazilian agencies, CNPq under grant 310038/2019-7, CAPES (88881.068485/2014) and FAPESP
(LLAMA Project no. 11/51676-9). GL acknowledges the partial financial support from  CNPq (grant 316844/2021-7) and FAPESP (grant 2022/02341-9).

\bibliography{references}
\end{document}